\documentclass[conference]{IEEEtran}
\usepackage[letterpaper, left=0.625in, right=0.625in, bottom=0.99in, top=0.7in]{geometry}
\IEEEoverridecommandlockouts
\usepackage{cite}
\usepackage{amsmath,amssymb,amsfonts}
\usepackage{algorithm}
\usepackage{algpseudocode}
\usepackage{graphicx}
\usepackage{textcomp}
\usepackage{xcolor}
\usepackage{bm}
\usepackage{booktabs}
\usepackage{array}
\usepackage{multirow}
\usepackage{comment}
\usepackage{tikz}
\usetikzlibrary{positioning, arrows.meta}
\def\BibTeX{{\rm B\kern-.05em{\sc i\kern-.025em b}\kern-.08em
    T\kern-.1667em\lower.7ex\hbox{E}\kern-.125emX}}

\begin{document}

\title{Beyond Binary Encoding: Qudit-Based QAOA Framework for ML Detection in $M$-QAM MIMO Systems}

\author{
\IEEEauthorblockN{Chiran Siriwardena\IEEEauthorrefmark{1}, Chanula Luckshan\IEEEauthorrefmark{1}, Samiru Gayan\IEEEauthorrefmark{1}, Hazer Inaltekin\IEEEauthorrefmark{3}, and Wonjae Shin\IEEEauthorrefmark{4}}

\IEEEauthorblockA{\IEEEauthorrefmark{1}Department of Electronic and Telecommunication Engineering, University of Moratuwa, Katubedda, Sri Lanka.}

\IEEEauthorblockA{\IEEEauthorrefmark{3}School of Engineering, Macquarie University, North Ryde, NSW 2109, Australia.}

\IEEEauthorblockA{\IEEEauthorrefmark{4}School of Electrical Engineering, Korea University, Seoul 02841, South Korea.}

\IEEEauthorblockA{Email:\{siriwardhanacchper.23, luckshangwcm.20, samirug\}@uom.lk, hazer.inaltekin@mq.edu.au, wjshin@korea.ac.kr}

\thanks{}
}

\maketitle

\begin{abstract}
Maximum likelihood (ML) detection in massive multiple-input multiple-output (MIMO) systems is an NP-hard combinatorial problem whose complexity grows exponentially with the number of transmit antennas and the modulation order. The Quantum Approximate Optimization Algorithm (QAOA) has recently emerged as a promising near-term quantum heuristic for such problems; however, existing qubit-based formulations require $\log_2 M$ qubits per $M$-QAM symbol, inflating both circuit width and depth. In this paper, we propose a \emph{qudit}-based QAOA framework that encodes each $M$-QAM symbol directly into a pair of $d$-dimensional quantum states, where $d=\sqrt{M}$. We derive a Hermitian cost Hamiltonian via a real-imaginary decomposition of the constellation, and introduce a generalized cyclic-shift mixer that ensures ergodic exploration of the search space without binary encoding. The proposed formulation reduces the number of quantum carriers from $\mathcal{O}(N_t \log_2 M)$ to $\mathcal{O}(N_t)$ and the Hamiltonian interaction complexity from $\mathcal{O}((N_t \log_2 M)^2)$ to $\mathcal{O}(N_t^2)$, while preserving the geometric structure of the constellation. Bit-error-rate (BER) simulations for $2\times 2$ MIMO systems employing 4-QAM and 16-QAM over Rayleigh fading channels, with circuit depth $p=3$, demonstrate that the proposed detector closely approximates the performance of the optimal classical maximum-likelihood (ML) detector, while consistently outperforming the conventional minimum mean square error (MMSE) detector.
\end{abstract}


\section{Introduction}

The relentless growth in wireless data demand has placed massive MIMO at the heart of the physical-layer design of 5G-Advanced and 6G systems. While linear receivers such as zero-forcing (ZF) and minimum mean-square error (MMSE) detection are computationally tractable, they fall short of the optimal bit error rate (BER) achievable by maximum likelihood (ML) detection. Unfortunately, ML detection is equivalent to the Closest Vector Problem (CVP) on a lattice and is therefore NP-hard~\cite{verdu89-4}: the search space scales as $M^{N_t}$, where $M$ is the modulation order and $N_t$ the number of transmit antennas. Sphere decoders, semidefinite-relaxation methods, and tabu search have all been proposed to bridge this gap, yet none scale gracefully to the hundreds of antennas and high-order constellations envisioned for 6G systems.

Quantum computing offers a fundamentally different attack on this computational bottleneck. Among near-term quantum heuristics, the Quantum Approximate Optimization Algorithm (QAOA)~\cite{farhi14-1} has attracted particular attention because its hybrid quantum–classical structure suits the limited coherence times and gate fidelities of Noisy Intermediate-Scale Quantum (NISQ) devices. Recent works~\cite{Cui22-2, cui22-5, Liu24-7, Burhan25-8, liu25-9, urgelles24-10} have shown that ML detection can be cast as a Quadratic Unconstrained Binary Optimization (QUBO) problem and embedded into a problem Hamiltonian whose ground state encodes the ML solution.

A central limitation of all of these prior works, however, lies in their reliance on \emph{qubit}-based encodings. To represent a single $M$-QAM symbol, $\log_2 M$ qubits are required, so a MIMO system with $N_t$ transmit antennas demands $N_t \log_2 M$ qubits. The binary decomposition is more than a bookkeeping overhead: it inflates the number of pairwise interaction terms in the Hamiltonian by a factor of $(\log_2 M)^2$, deepens the QAOA circuit, distorts the geometric structure of the constellation (Hamming-adjacent codewords are not generally Euclidean-adjacent symbols), and amplifies decoherence. For high-order constellations such as 256-QAM and 1024-QAM, the resulting circuits are prohibitive on any foreseeable NISQ hardware.

In this paper, we argue that the natural quantum representation of an $M$-QAM symbol is not a string of qubits but a single high-dimensional \emph{qudit}. Building on this observation, we make the following contributions:

\begin{enumerate}
\item We derive a qudit-based Hermitian cost Hamiltonian for ML detection that directly encodes $M$-QAM symbols into two $d$-dimensional qudit states ($d=\sqrt{M}$), corresponding to the in-phase and quadrature components via a real-imaginary decomposition of the constellation. This preserves the true Euclidean geometry of the symbols and extends to higher-order modulation.
\item We construct a generalized cyclic-shift mixer Hamiltonian that ensures ergodic exploration of the qudit search space and admits an efficient implementation using generalized $RZ_d(\theta)$ phase rotations.

\item We present a comparative complexity analysis showing that the proposed formulation significantly reduces quantum carrier requirements compared to qubit-based general QAOA ~\cite{cui22-5} from $\mathcal{O}(N_t \log_2 M)$ to $\mathcal{O}(N_t)$, while compared to naive binary QAOA, the interaction complexity reduces from $\mathcal{O}((N_t \log_2 M)^2)$ to $\mathcal{O}(N_t^2)$. 

\item We validate the framework via QuTiP simulations of a $2\times 2$ MIMO 4-QAM and $2\times 2$ MIMO 16-QAM system over a Rayleigh fading channel, showing that the proposed detector closely tracks classical ML and consistently outperforms MMSE.
\end{enumerate}

The remainder of the paper is organized as follows. Section~\ref{sec:related} reviews related work. Section~\ref{sec:system} formulates the ML detection problem and its real-valued equivalent. Section~\ref{sec:proposed} introduces the qudit-based QAOA framework, its gate-level implementation, and pseudocode. Section~\ref{sec:comparison} compares the proposed approach with qubit-based QAOA and presents simulation results. Section~\ref{sec:conclusion} concludes. Proofs are deferred to the appendix.

\section{Related Work}\label{sec:related}

The hardness of ML detection in MIMO systems is established in~\cite{verdu89-4}, and closed-form performance analyses for fading channels are presented in~\cite{Zhu02-3}. Sphere decoding and lattice-reduction techniques approach ML performance while maintaining reduced average complexity, but retain exponential worst-case complexity, motivating the search for quantum alternatives. 

Cui \emph{et al.}~\cite{Cui22-2} were among the first to formulate ML detection as a QUBO and apply QAOA, demonstrating performance close to classical ML for BPSK and QPSK at modest depth. Their follow-up~\cite{cui22-5} extends the construction to higher-order modulation through binary (Gray) encoding, where each symbol is represented using $\log_2(M)$ qubits. This leads to high-degree polynomial Hamiltonians with multi-body Pauli-Z interactions that must be reduced to two-local form via quadratization, resulting in significant overhead for higher-order QAM. Liu \emph{et al.}~\cite{Liu24-7} address the massive-MIMO regime through warm-start initializations, while G\"{u}lbahar~\cite{Burhan25-8} proposes a recursive QAOA with majority-voting and cost-restricted sampling to improve solution quality for large $N_t$. The recent shallow-compiled QAOA framework~\cite{liu25-9} reduces compilation overhead through pre-trained parameter schedules, and Urgelles \emph{et al.}~\cite{urgelles24-10} extend the QAOA-ML approach to MIMO-NOMA systems for 6G. All of these works, however, share the qubit-based encoding bottleneck identified above.

Recent work such as Kim \emph{et al.}~\cite{kim2025xresq} employs quantum annealing and QUBO formulations for ML MIMO detection, achieving hardware-level parallelism but relying on binary encodings. Beyond annealing, qudit-based generalizations of QAOA have been explored for inherently $d$-ary problems such as graph coloring~\cite{Deller23-6}, demonstrating advantages in circuit width and depth, while Pudda \emph{et al.}~\cite{pudda2024-13} provide a systematic gate-level framework for qudit decompositions. However, to the best of our knowledge, no prior work has applied qudit-based QAOA to ML MIMO detection or developed a Hamiltonian specifically tailored to $M$-QAM constellations. The proposed framework fills this gap and is the first to leverage the natural alignment between $\sqrt{M}$-PAM-per-axis structure and $d$-dimensional qudit registers. Whereas prior QAOA-MIMO works trade qubit count and circuit depth against detection performance through algorithmic refinements (warm-starting, recursion, sampling), our work attacks the underlying \emph{representation} bottleneck. The qudit framework is complementary to and composable with these techniques: warm-starting and recursive QAOA can be applied atop the proposed Hamiltonian to further reduce the variational depth.

\section{System Model and Problem Formulation}\label{sec:system}

\subsection{MIMO System Model}

We consider a point-to-point MIMO link with $N_t$ transmit antennas and $N_r$ receive antennas. The complex baseband received vector $\mathbf{y}\in\mathbb{C}^{N_r}$ is
\begin{equation}
\mathbf{y} = \mathbf{H}\mathbf{x} + \mathbf{n},
\label{eq:mimo}
\end{equation}
where $\mathbf{H}\in\mathbb{C}^{N_r\times N_t}$ is the channel matrix, $\mathbf{x}\in\mathcal{A}^{N_t}$ is the transmitted vector drawn from an $M$-QAM constellation $\mathcal{A}$, and $\mathbf{n}\sim\mathcal{CN}(\mathbf{0},\sigma^2\mathbf{I})$ is additive white Gaussian noise. Assuming perfect channel state information at the receiver, the ML detector solves
\begin{equation}
\hat{\mathbf{x}}_{\mathrm{ML}} = \arg\min_{\mathbf{x}\in\mathcal{A}^{N_t}} \|\mathbf{y}-\mathbf{H}\mathbf{x}\|_2^2.
\label{eq:ml}
\end{equation}
Expanding~(\ref{eq:ml}) and dropping the constant term $\mathbf{y}^\dagger\mathbf{y}$ yields the equivalent quadratic objective
\begin{equation}
f(\mathbf{x}) = \mathbf{x}^\dagger\mathbf{A}\mathbf{x} - 2\,\Re\{\mathbf{b}^\dagger\mathbf{x}\},
\quad \mathbf{A}=\mathbf{H}^\dagger\mathbf{H},\ \mathbf{b}=\mathbf{H}^\dagger\mathbf{y}.
\label{eq:cost-complex}
\end{equation}

\subsection{Real-Valued Equivalent Formulation}

To facilitate the qudit encoding introduced in Section~\ref{sec:proposed}, we recast~(\ref{eq:cost-complex}) in real-valued form. Define
\begin{equation}
\tilde{\mathbf{y}}=\begin{bmatrix}\Re(\mathbf{y})\\ \Im(\mathbf{y})\end{bmatrix},\quad
\tilde{\mathbf{x}}=\begin{bmatrix}\Re(\mathbf{x})\\ \Im(\mathbf{x})\end{bmatrix},\quad
\tilde{\mathbf{H}}=\begin{bmatrix}\Re(\mathbf{H}) & -\Im(\mathbf{H})\\ \Im(\mathbf{H}) & \Re(\mathbf{H})\end{bmatrix}. \nonumber
\end{equation}
With $\tilde{\mathbf{A}}=\tilde{\mathbf{H}}^T\tilde{\mathbf{H}}\in\mathbb{R}^{2N_t\times 2N_t}$ and $\tilde{\mathbf{b}}=\tilde{\mathbf{H}}^T\tilde{\mathbf{y}}\in\mathbb{R}^{2N_t}$, problem~(\ref{eq:ml}) becomes
\begin{equation}
\min_{\tilde{\mathbf{x}}\in\tilde{\mathcal{A}}^{2N_t}} \tilde{\mathbf{x}}^T\tilde{\mathbf{A}}\tilde{\mathbf{x}} - 2\tilde{\mathbf{b}}^T\tilde{\mathbf{x}}.
\label{eq:cost-real}
\end{equation}

\noindent
Let us define $\mathbf{R} = \Re(\mathbf{H})$ and $\mathbf{I} = \Im(\mathbf{H})$. 
Then, the real-valued Gram matrix $\tilde{\mathbf{A}}$ and vector $\tilde{\mathbf{b}}$ are given by
\begin{equation}
\tilde{\mathbf{A}} =
\begin{bmatrix}
\mathbf{R}^T \mathbf{R} + \mathbf{I}^T \mathbf{I} & -\mathbf{R}^T \mathbf{I} + \mathbf{I}^T \mathbf{R} \\
-\mathbf{I}^T \mathbf{R} + \mathbf{R}^T \mathbf{I} & \mathbf{R}^T \mathbf{R} + \mathbf{I}^T \mathbf{I}
\end{bmatrix}
=
\begin{bmatrix}
\mathbf{G} & \mathbf{C}\\
-\mathbf{C} & \mathbf{G}
\end{bmatrix}.
\end{equation}

\begin{equation}
\tilde{\mathbf{b}} =
\begin{bmatrix}
\mathbf{R}^T \Re(\mathbf{y}) + \mathbf{I}^T \Im(\mathbf{y}) & \\
-\mathbf{I}^T \Re(\mathbf{y}) + \mathbf{R}^T \Im{(\textbf{y})} & 
\end{bmatrix}
=
\begin{bmatrix}
\textbf{s}\\
\textbf{t}
\end{bmatrix}.
\end{equation}

Then the quadratic term can be expanded as:

\begin{align}
\tilde{\mathbf{x}}^T \tilde{\mathbf{A}} \tilde{\mathbf{x}} &= \Re(\mathbf{x}^T) \mathbf{G} \Re(\mathbf{x}) + \Re(\mathbf{x}^T) \mathbf{C} \Im(\mathbf{x}) - \Im(\mathbf{x}^T)\mathbf{C} \Re(\mathbf{x}) \nonumber \\
&+ \Im(\mathbf{x}^T) \mathbf{G} \Im(\mathbf{x})
\end{align}

\noindent With the property
\begin{equation}
\begin{aligned}
\mathbf{C}^T = (-\mathbf{R}^T \mathbf{I} + \mathbf{I}^T \mathbf{R})^T = -\mathbf{I}^T \mathbf{R} + \mathbf{R}^T \mathbf{I} = -\mathbf{C}
\end{aligned}
\end{equation}

\noindent The objective function can be expanded with cross term $\Re(\mathbf{x}) \Im(\mathbf{x})$ and with the definition of $\tilde{\mathbf{x}}^R=\Re(\mathbf{x})$, $\tilde{\mathbf{x}}^I=\Im(\mathbf{x})$
\begin{equation}
f(\tilde{\mathbf{x}}) =
(\tilde{\mathbf{x}}^R)^T \mathbf{G} \tilde{\mathbf{x}}^R
+ (\tilde{\mathbf{x}}^I)^T \mathbf{G} \tilde{\mathbf{x}}^I
+ 2(\tilde{\mathbf{x}}^R)^T \mathbf{C} \tilde{\mathbf{x}}^I
- 2\mathbf{s}^T\tilde{\mathbf{x}}^R
- 2\mathbf{t}^T \tilde{\mathbf{x}}^I
\end{equation}

Therefore, the objective can be written explicitly as
\begin{align}
f(\tilde{\mathbf{x}}) &=
\sum_{i=1}^{N_t}\sum_{j=1}^{N_t}
G_{ij} \tilde{x}_i^R \tilde{x}_j^R
+ G_{ij} \tilde{x}_i^I \tilde{x}_j^I 
+ 2 C_{ij} \tilde{x}_i^R \tilde{x}_j^I \nonumber\\
&- 2\sum_{i=1}^{N_t} \Big(s_i \tilde{x}_i^R+ t_i \tilde{x}_i^I\Big)
\label{eq:obj_function}
\end{align}

\section{Proposed Qudit-Based QAOA Framework}\label{sec:proposed}

\subsection{QAOA Preliminaries}

QAOA prepares a parameterized trial state by alternately applying a problem (cost) Hamiltonian $\hat{H}_f$, which is diagonal in the computational basis, and a non-diagonal mixer Hamiltonian $\hat{H}_B$. Starting from the uniform superposition $|\psi_0\rangle$, the depth-$p$ QAOA state is
\begin{equation}
|\psi_p(\bm{\gamma},\bm{\beta})\rangle = \prod_{k=1}^{p} e^{-i\beta_k \hat{H}_B}\,e^{-i\gamma_k \hat{H}_f}\, |\psi_0\rangle.
\label{eq:qaoa-state}
\end{equation}
The variational parameters $(\bm{\gamma},\bm{\beta})$ are tuned by a classical outer loop to minimize the cost expectation $F_p(\bm{\gamma},\bm{\beta}) = \langle\psi_p|\hat{H}_f|\psi_p\rangle$.
Sampling $|\psi_p(\bm{\gamma}^*,\bm{\beta}^*)\rangle$ then yields high-quality solutions to the underlying combinatorial problem.

\subsection{Qubit-Based Baseline for BPSK}

For BPSK, the binary mapping $s_k=1-2z_k\in\{-1,+1\}$ converts the cost~(\ref{eq:cost-real}) into a QUBO whose Hamiltonian is
\begin{equation}
\hat{H}_{f,\mathrm{BPSK}} = \sum_{i<j}^{N_t} 2\tilde{A}_{ij}\hat{\sigma}_z^i \hat{\sigma}_z^j - \sum_{i=1}^{N_t} 2\tilde{b}_i \hat{\sigma}_z^i,
\label{eq:H-bpsk}
\end{equation}
with the standard Pauli-$X$ mixer $\hat{H}_B=-\sum_i \hat{\sigma}_x^i$. As shown in Proposition~\ref{prop:bpsk}, the eigenstates of~(\ref{eq:H-bpsk}) are in one-to-one correspondence with BPSK candidate vectors. Extending this construction to $M$-QAM via binary encoding requires $\log_2 M$ qubits per symbol. Although naive binary formulations incur a quadratic increase in the number of interaction terms, more advanced approaches \cite{cui22-5} exploit constellation structure to achieve linear scaling in $\log_2 M$. However, such methods still rely on binary representations, leading to higher quantum-carrier requirements than the proposed qudit-based formulation.

\subsection{Qudit Encoding of M-QAM Symbols}

Let $d=\sqrt{M}$. We represent each transmitted $M$-QAM symbol $x_j$ by a tensor product of two qudits: one encoding the in-phase level and one encoding the quadrature level. Each qudit lives in $\mathbb{C}^d$ and has computational basis $\{|0\rangle,|1\rangle,\ldots,|d-1\rangle\}$. The two-qudit symbol state is
\begin{equation}
|x_j\rangle = |k\rangle\otimes|l\rangle = |kl\rangle, \quad k,l\in\{0,\ldots,d-1\},
\end{equation}
and the joint transmit state for $N_t$ antennas is $|\mathbf{x}\rangle = \bigotimes_{j=1}^{N_t} |x_j\rangle \in (\mathbb{C}^d\otimes\mathbb{C}^d)^{\otimes N_t} \cong \mathbb{C}^{M^{N_t}}.$
The bijection between basis states and constellation points is
\begin{equation}
g(z_I, z_Q) = \big(2 z_I-(d-1)\big) + j\big(2 z_Q-(d-1)\big),
\label{eq:bijection}
\end{equation}
which preserves the constellation's rectangular geometry. Crucially, adjacent qudit basis states map to Euclidean-adjacent constellation points, so the cyclic mixer introduced below induces local moves in signal space.

\subsection{Qudit Cost Hamiltonian}

We define two diagonal Hermitian operators that act on a single two-qudit symbol register and return its real and imaginary components:
\begin{align}
\hat{x}_R &= \sum_{k,l=0}^{d-1} \Re(a_{kl})\,|kl\rangle\langle kl|,\label{eq:xR}\\
\hat{x}_I &= \sum_{k,l=0}^{d-1} \Im(a_{kl})\,|kl\rangle\langle kl|,\label{eq:xI}
\end{align}
where $a_{kl}=g(k,l)$ from~(\ref{eq:bijection}). Both operators are Hermitian (the diagonal entries are real), so they correspond to physically realizable observables. Using copies $\hat{x}_i^R, \hat{x}_i^I$ acting on the $i$-th symbol register, we substitute~(\ref{eq:xR})--(\ref{eq:xI}) into~(\ref{eq:obj_function}) to obtain the proposed cost Hamiltonian
\begin{align}\label{eq:H-mqam}
\hat{H}_{f,\mathrm{M-QAM}} =\;&
\sum_{i,j=1}^{N_t} G_{ij} \hat{x}_i^R\hat{x}_j^R + G_{ij} \hat{x}_i^I\hat{x}_j^I + 2C_{ij} \hat{x}_i^R\hat{x}_j^I \nonumber \\
&- 2\sum_{i=1}^{N_t}\big(s_i\hat{x}_i^R + t_i \hat{x}_i^I\big)
\end{align}

By construction (see Proposition~\ref{prop:mqam}), every computational basis state $|\mathbf{x}\rangle$ is an eigenstate of~(\ref{eq:H-mqam}) with eigenvalue equal to the classical cost~(\ref{eq:obj_function}) evaluated at the corresponding constellation vector. Hence, the ground state of $\hat{H}_{f,\mathrm{M-QAM}}$ encodes the ML solution.

\subsection{Qudit Mixer Hamiltonian and Initial State}

The role of the mixer is to redistribute amplitude between basis states so that the QAOA evolution can escape local minima. In the qubit setting, this is the Pauli-$X$ operator; in the qudit setting, we adopt the generalized cyclic-shift operator
\begin{equation}
\hat{X}|z\rangle = |(z+1)\bmod d\rangle,
\end{equation}
and define the mixer as a sum over all $2N_t$ qudits in the system,
\begin{equation}
\hat{H}_B = -\sum_{k=1}^{2N_t}\!\big(\hat{X}_k + \hat{X}_k^{\dagger}\big).
\label{eq:mixer}
\end{equation}
$\hat{H}_B$ is Hermitian and generates transitions within the cyclic group $\mathbb{Z}_d$ on each qudit register. Repeated application of $\hat{H}_B$ connects every pair of computational basis states, ensuring ergodic exploration of the search space. The initial state is the uniform superposition
\begin{equation}
|\psi_0\rangle = \frac{1}{\sqrt{d^N}}\sum_{z=0}^{d^N-1}|z\rangle, \quad N=2N_t,
\end{equation}
so all $M^{N_t}$ candidate symbol vectors are explored with equal probability at $p=0$.

\subsection{Gate-Level Implementation}

Because $\hat{x}_i^R$ and $\hat{x}_i^I$ are diagonal, the cost unitary $e^{-i\gamma\hat{H}_{f,\mathrm{M-QAM}}}$ factorizes into a product of single- and two-qudit phase rotations. For a single diagonal operator $\hat{D}=\sum_k d_k|k\rangle\langle k|$,
\begin{equation}
e^{-i\gamma\hat{D}} = \mathrm{diag}\big(e^{-i\gamma d_0},\ldots,e^{-i\gamma d_{d-1}}\big),
\label{eq:diag-exp}
\end{equation}
which is precisely the action of a generalized $RZ_d(\gamma)$ gate. Equivalently, it can be expressed as a product of single-level phase gates $P_d^k(\theta)
=\mathrm{diag}(1,\ldots,e^{i\theta},\ldots,1)$, $RZ_d(\gamma) =\prod_{k=0}^{d-1}P_d^k(-\gamma d_k).$

The operators $-2s_i\hat{x}_i^{R}$ and $-2t_i\hat{x}_i^{I}$ exponentiate into single-ququart ($d$ = 4 qudit) $RZ_d$ gates with rotation parameters proportional to $\gamma s_i$ or $\gamma t_i$. The interaction $G_{ij}\hat{x}_i^R\hat{x}_j^R$ is implemented by a controlled $RZ_d$ between the two ququarts, denoted $C\text{-}RZ_d$, whose effect is
\begin{equation}
C\text{-}RZ_d(\gamma G_{ij}):\; |k\rangle|l\rangle \mapsto e^{-i\gamma G_{ij}\,d_k d_l}|k\rangle|l\rangle.
\end{equation}
The diagonal squared term $G_{ii}(\hat{x}_i^R)^2$ is itself diagonal and therefore reduces to a single-ququart $RZ_d$ rotation with angles proportional to $d_k^2$. The cyclic-shift operator $\hat{X}$ is diagonalized by the qudit Fourier transform $F_d$, yielding $\hat{X}=F_d^\dagger Z_d F_d$, where $Z_d=\mathrm{diag}(1,\omega,\omega^2,\ldots,\omega^{d-1})$ with $\omega=e^{2\pi i/d}$. The mixer unitary $e^{-i\beta\hat{H}_B}$ on each ququart therefore decomposes as
\begin{equation}
e^{i\beta(\hat{X}+\hat{X}^\dagger)} = F_d^\dagger\, e^{i\beta(Z_d + Z_d^\dagger)}\, F_d,
\label{eq:mixer-impl}
\end{equation}
which is a sandwich of two qudit Fourier transforms around a diagonal phase rotation -- a primitive available natively in trapped-ion and photonic qudit platforms~\cite{pudda2024-13}. On hardware that does not support qudit gates natively, each generalized $RZ_d$ rotation decomposes into $\mathcal{O}(d)$ single-level phase gates, and each $C\text{-}RZ_d$ requires $\mathcal{O}(d^2)$ two-level controlled rotations. The qudit Fourier transform $F_d$ likewise decomposes into $\mathcal{O}(d^2)$ two-level operations~\cite{pudda2024-13}. 

\subsection{QAOA Algorithm}

\begin{figure*}[t]
\centering
\includegraphics[width=0.52\textwidth]{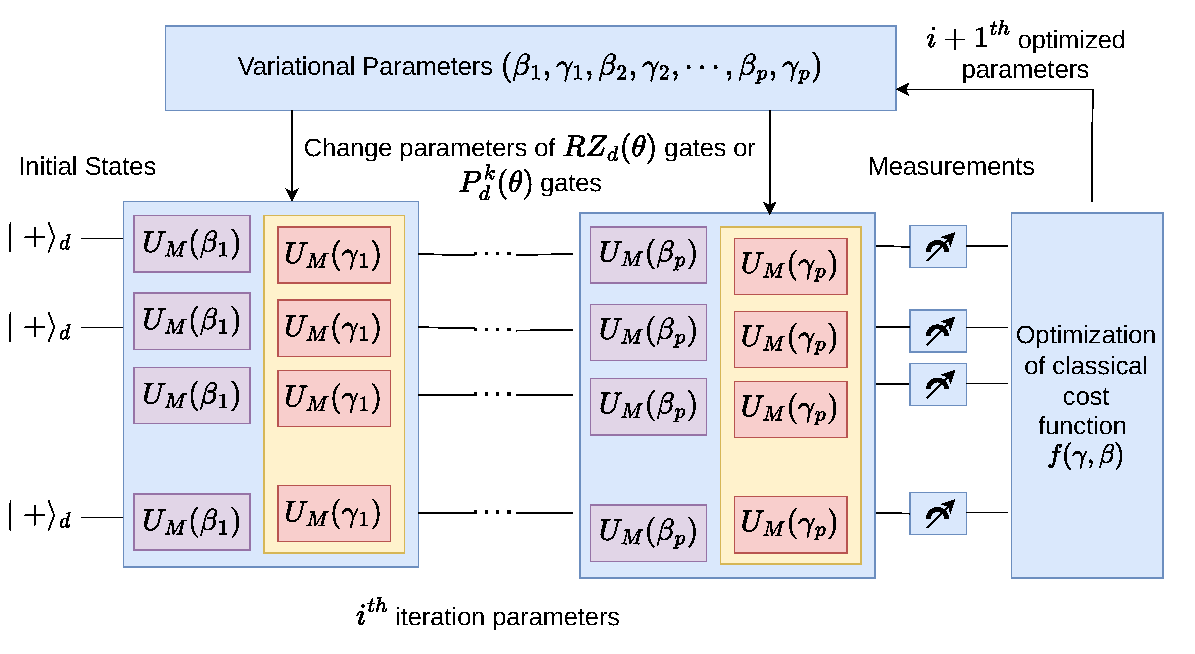}
\caption{Proposed qudit-based QAOA framework. Generalized $RZ_d(\theta)$ and $P_d^k(\theta)$ gates implement the cost and mixer unitaries; classical optimization updates $(\bm{\gamma},\bm{\beta})$ between iterations.}
\label{fig:framework}
\end{figure*}

Algorithm~\ref{alg:qaoa} summarizes the proposed qudit-QAOA detector. The classical optimizer can be any gradient-free routine; in our simulations, we use COBYLA with multi-start initialization, since QAOA cost landscapes are known to be non-convex with shallow local minima at small $p$~\cite{zhou20-11}. Parameter initialization is drawn from the linear-ramp heuristic, $\gamma_k = (k/p)\Delta\gamma$ and $\beta_k=(1-k/p)\Delta\beta$ with $\Delta\gamma,\Delta\beta\in[0,\pi]$, which is known to converge faster than uniformly random starts. The full pipeline is depicted in Fig.~\ref{fig:framework}.

\begin{algorithm}[t]
\caption{Qudit-QAOA ML Detector for $M$-QAM MIMO}
\label{alg:qaoa}
\begin{algorithmic}[1]
\Require Channel $\mathbf{H}$, received vector $\mathbf{y}$, modulation order $M$, depth $p$, classical optimizer \textsc{Opt}
\Ensure Estimated symbol vector $\hat{\mathbf{x}}$
\State $d \gets \sqrt{M}$
\State Form real-valued $\tilde{\mathbf{H}},\tilde{\mathbf{y}}$ and compute $\tilde{\mathbf{A}},\tilde{\mathbf{b}}$ \Comment{Eq.~(\ref{eq:cost-real})}
\State Construct diagonal operators $\hat{x}_i^R,\hat{x}_i^I$ on $\mathbb{C}^{d^2}$ \Comment{Eqs.~(\ref{eq:xR})--(\ref{eq:xI})}
\State Build $\hat{H}_{f,\mathrm{M-QAM}}$ from~(\ref{eq:H-mqam}) and $\hat{H}_B$ from~(\ref{eq:mixer})
\State Initialize $(\bm{\gamma}_0,\bm{\beta}_0)$ via linear-ramp heuristic
\State Prepare $|\psi_0\rangle = d^{-N_t}\sum_z|z\rangle$
\Repeat
  \State Apply $|\psi_p\rangle = \prod_{k=1}^{p} e^{-i\beta_k\hat{H}_B} e^{-i\gamma_k\hat{H}_{f,\mathrm{M-QAM}}}|\psi_0\rangle$
  \State Estimate $F_p(\bm{\gamma},\bm{\beta}) = \langle\psi_p|\hat{H}_{f,\mathrm{M-QAM}}|\psi_p\rangle$
  \State Update $(\bm{\gamma},\bm{\beta}) \gets \textsc{Opt}(F_p)$
\Until convergence or budget exhausted
\State Sample $|\psi_p(\bm{\gamma}^*,\bm{\beta}^*)\rangle$ to obtain $|\hat{\mathbf{x}}\rangle$
\State Return $\hat{\mathbf{x}}$ via inverse map of~(\ref{eq:bijection})
\end{algorithmic}
\end{algorithm}

\section{Complexity Analysis and Numerical Results}\label{sec:comparison}

\subsection{Quantum Resource Comparison}

\begin{table}[t]
\centering
\caption{Resource comparison of qubit-based and qudit-based QAOA for $M$-QAM ML detection.}
\label{tab:resources}
\begin{tabular}{@{}lcc@{}}
\toprule
\textbf{Metric} & \textbf{Qubit-based} & \textbf{Qudit-based (proposed)} \\
\midrule
Carriers per symbol & $\log_2 M$ qubits & $2$ qudits ($d=\sqrt{M}$) \\
Total carriers$^{1}$ & $N_t\log_2 M$ & $2N_t$ \\
Circuit Depth$^{1}$ & $\mathcal{O}(p \cdot T_k \cdot k)$ & $\mathcal{O}\big(p \cdot N_t\big)$ \\
Interactions$^{2}$ & $\mathcal{O}((N_t\log_2(M))^2)$ & $\mathcal{O}\big(N^2_t\big)$ \\
Constellation geometry & distorted & preserved \\
\bottomrule
\multicolumn{3}{p{0.95\columnwidth}}{
\scriptsize
$^{1}$ Compared to SAT-based qubit QAOA method of \cite{cui22-5}. \newline $^{2}$ Compared to naive dense-QUBO binary encoding.
}
\end{tabular}
\end{table}

Table~\ref{tab:resources} summarizes the comparison. The proposed method reduces the number of quantum carriers from $\mathcal{O}(N_t \log_2 M)$ to $\mathcal{O}(N_t)$ by employing qudits, and achieves improved circuit-depth scaling compared to the SAT-based approach in~\cite{cui22-5}. The corresponding reduction ratio is $\eta = 1 - \frac{2}{\log_2 M},$
which increases with $M$ and approaches $80\%$ for 1024-QAM. The proposed Hamiltonian exhibits interaction complexity $\mathcal{O}(N_t^2)$, depending only on $N_t$. In contrast, naive binary QAOA scales as $\mathcal{O}((N_t \log_2 M)^2)$, yielding a complexity reduction by a factor of $\mathcal{O}((\log_2 M)^2)$. From a circuit depth perspective, the proposed M-QAM qudit Hamiltonian offers a more favorable structure than the SAT-based formulation in~\cite{cui22-5}. The Hamiltonian in~(\ref{eq:H-mqam}) contains $O(N_t^2)$ two-body and $O(N_t)$ single-body terms. Since all operators are diagonal and mutually commuting, the unitary evolution can be implemented without Trotterization. Under ideal all-to-all connectivity, all terms can be applied in parallel, yielding constant depth per QAOA layer. With limited connectivity, edge-coloring arguments imply a depth scaling of $O(N_t)$ per layer. Including the mixer, the overall circuit depth scales as $D_{\mathrm{M-QAM}} = \mathcal{O}(p\,N_t)$. In contrast, the SAT-based approach in~\cite{cui22-5} maps ML detection to a weighted $N$-SAT problem, yielding a multi linear polynomial of degree $k=\mathcal{O}(\log_2 M)$ with corresponding $k$-local Hamiltonian terms:
\begin{equation}
H_f = \sum_{S \subseteq N} g_S \prod_{i \in S} \sigma_z^{(i)}.
\end{equation}
\noindent Since near-term quantum hardware natively supports only two-local interactions, these higher-order terms must be implemented via decomposition into sequences of two-qubit gates, potentially introducing auxiliary qubits depending on the chosen reduction method. Each $k$-local term can be realized using $\mathcal{O}(k)$ two-qubit gates, leading to a circuit depth per QAOA layer that scales as $\mathcal{O}(T_k \cdot k)$, where $T_k$ denotes the number of nonzero $k$-local terms. Accordingly, for a depth-$p$ QAOA circuit, the overall depth scales as $D_{\mathrm{SAT}} = \mathcal{O}(p \cdot T_k \cdot k)$. Although many higher-order terms may vanish due to constellation structure and Gray mapping, the remaining multi-qubit interactions and their decomposition introduce nontrivial overhead. In contrast, by avoiding high-order interactions and preserving a quadratic structure, the proposed Hamiltonian enables more hardware-efficient QAOA implementations.

\subsection{Scalability}

For high-order constellations, the advantage of qudit encoding compounds. The qubit circuit width scales as $\mathcal{O}(N_t\log_2 M)$, whereas the qudit width scales as $\mathcal{O}(N_t)$, independent of $M$. This linear scaling in $N_t$ alone is critical for envisioned 6G use cases such as 256-QAM and 1024-QAM transmission over massive MIMO arrays.

\subsection{Simulation Setup}

We simulate $2\times 2$ MIMO systems with 4-QAM and 16-QAM over an i.i.d.\ Rayleigh fading channel using the QuTiP framework. The QAOA depth is set to $p=3$, and classical optimization of $(\bm{\gamma},\bm{\beta})$ uses COBYLA with a multi-start initialization. SNR ranges from $-4$ dB to $0$ dB, where the BER is averaged over $5{,}000$ independent channel realizations. For $0$ dB to $8$ dB, the BER is averaged over $8{,}000$ independent channel realizations. Finally, in the high-SNR regime from $8$ dB to $12$ dB, the BER is averaged over $13{,}000$ independent channel realizations. We compare three detectors: (i) classical exhaustive ML, (ii) MMSE, and (iii) the proposed qudit-QAOA detector.

\subsection{BER Performance}

\begin{figure}
    \centering
    \includegraphics[width=0.38\textwidth]{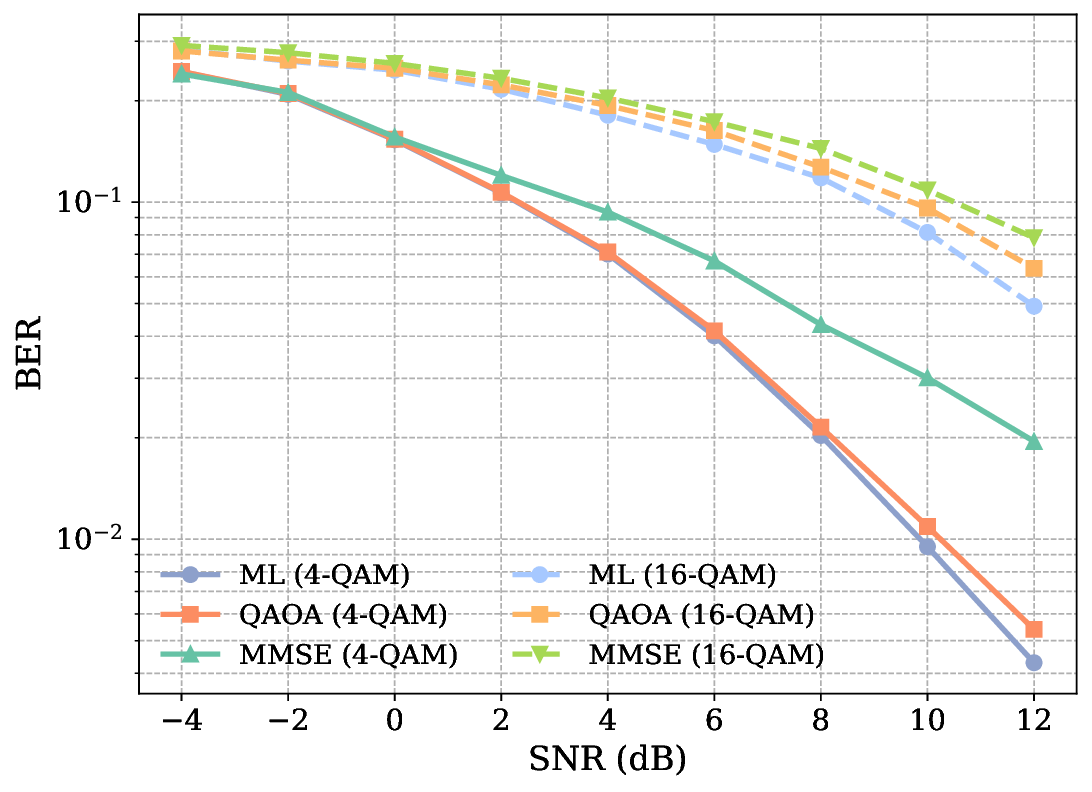}
    \caption{BER vs.\ SNR over Rayleigh fading channels.}
    \label{fig:ber_4QAM_16QAM}
\end{figure}

Figure~\ref{fig:ber_4QAM_16QAM} reports BER as a function of SNR over the range $-4$ to $12$~dB. Three observations stand out. First, the proposed qudit-QAOA detector closely tracks the optimal ML curve at low and moderate SNR, where the noise dominates the residual variational error. Second, qudit-QAOA consistently outperforms MMSE across the full SNR range for 4-QAM, confirming that even at modest depth $p=3$, the variational ansatz captures enough of the cost landscape to surpass linear receivers. Third, a small gap to the optimal ML persists at high SNR for both 4-QAM and 16-QAM, reflecting the inherently approximate nature of variational quantum optimization. We expect this gap to close further with deeper circuits or better classical optimizers, at the cost of more noise on real hardware, a trade-off discussed next.

\subsection{Effect of QAOA Depth and Circuit Complexity}

The performance of the QAOA ansatz improves with circuit depth $p$, as higher depths enable the capture of increasingly long-range correlations in the optimization landscape. At $p=1$, only a limited local structure is represented, while larger values of $p$ progressively approximate the ground state more accurately. Empirically, BER reductions of less than 0.5 dB were observed beyond $p = 3$ for the 4-QAM setting. However, on NISQ hardware, increasing $p$ also amplifies gate errors and decoherence effects~\cite{alam2019analysis-12}, which are further exacerbated in qudit systems due to higher leakage and more complex $d$-dimensional gate implementations. As a result, shallow-depth regimes (e.g., $p=1$--$2$) are typically optimal for near-term devices.

\subsection{Limitations and Outlook}

Two main limitations should be acknowledged. First, controllable qudit dimension is currently capped near $d\le 10$ on most platforms~\cite{pudda2024-13}, although photonic and trapped-ion systems are rapidly pushing this boundary; this would support up to 64-QAM with present hardware and 256-QAM in the near term. Second, multi-level control increases the number of possible transition pathways -- scaling as $\binom{d}{2}$ -- which complicates calibration and error mitigation, and leakage to non-computational levels becomes more pronounced. Future work will address noise-aware variational optimization, extension to larger MIMO arrays and higher-order modulation, integration with warm-start and recursive QAOA strategies~\cite{Liu24-7,Burhan25-8}, and evaluation on emerging photonic qudit testbeds.

\section{Conclusion}\label{sec:conclusion}

We presented a qudit-based QAOA framework for ML detection in MIMO systems with $M$-QAM modulation. By encoding each symbol into two $d$-dimensional qudits via a real--imaginary decomposition, the proposed cost Hamiltonian eliminates the binary-encoding overhead of qubit-based formulations and preserves the true Euclidean geometry of signals, and the proposed approach can be extended to higher-order M-QAM. A generalized cyclic-shift mixer ensures ergodic exploration of the qudit search space and admits an efficient implementation using native qudit phase rotations. A worked $2\times 2$ 16-QAM example confirms the predicted savings, and we provide complete pseudocode and gate-level decomposition to facilitate reproduction. Simulations of a $2\times 2$ MIMO 4-QAM system over Rayleigh fading confirm that the proposed detector closely tracks classical ML and consistently outperforms MMSE, even at modest QAOA depth. As qudit hardware matures, the framework is well-positioned to deliver scalable, near-optimal MIMO detection for high-order constellations envisioned for 6G systems.

\appendix

\subsection{Proof of Proposition 1 (BPSK)}\label{prop:bpsk}

\textbf{Proposition 1.} \emph{The eigenstates of $\hat{H}_{f,\mathrm{BPSK}}$ in~(\ref{eq:H-bpsk}) are computational basis states $|z\rangle$, with eigenvalues equal to the classical BPSK cost~(\ref{eq:cost-real}) under the bijection $s_k=1-2z_k$.}

\begin{IEEEproof}
The Pauli-$Z$ operator acts as $\hat{\sigma}_z^k|z_k\rangle = (1-2z_k)|z_k\rangle = s_k|z_k\rangle$. For a multi-qubit basis state $|\mathbf{z}\rangle$, $\hat{\sigma}_z^i|\mathbf{z}\rangle = s_i|\mathbf{z}\rangle$ and $\hat{\sigma}_z^i\hat{\sigma}_z^j|\mathbf{z}\rangle=s_is_j|\mathbf{z}\rangle$. Substituting into~(\ref{eq:H-bpsk}) gives
\begin{equation*}
\hat{H}_{f,\mathrm{BPSK}}|\mathbf{z}\rangle = \!\Big(\!\sum_{i<j}\!2\tilde{A}_{ij}s_is_j - \!\sum_i\! 2\tilde{b}_is_i\!\Big)|\mathbf{z}\rangle = f(\mathbf{s})|\mathbf{z}\rangle,
\end{equation*}
which establishes the claim. 
\end{IEEEproof}

\subsection{Proof of Proposition 2 (M-QAM)}\label{prop:mqam}

\textbf{Proposition 2.} \emph{The eigenstates of $\hat{H}_{f,\mathrm{M-QAM}}$ in~(\ref{eq:H-mqam}) are the computational basis states $|\mathbf{x}\rangle=\bigotimes_i|z_I^i z_Q^i\rangle$, with eigenvalues equal to the classical $M$-QAM cost~(\ref{eq:obj_function}) under the bijection $g$ in~(\ref{eq:bijection}).}

\begin{IEEEproof}
By construction~(\ref{eq:xR})--(\ref{eq:xI}), $\hat{x}_i^R$ and $\hat{x}_i^I$ are diagonal in the computational basis, so $\hat{x}_i^R|\mathbf{x}\rangle=\Re(a_i)|\mathbf{x}\rangle$ and $\hat{x}_i^I|\mathbf{x}\rangle=\Im(a_i)|\mathbf{x}\rangle$, where $a_i=g(z_I^i,z_Q^i)$. Products of diagonal operators remain diagonal, giving $\hat{x}_i^R\hat{x}_j^R|\mathbf{x}\rangle = \Re(a_i)\Re(a_j)|\mathbf{x}\rangle$ and similarly for the imaginary terms and cross terms. Substituting these eigenvalue equations into~(\ref{eq:H-mqam}) yields
\begin{equation*}
\hat{H}_{f, \mathrm{M-QAM}}|\mathbf{x}\rangle = f(\mathbf{a})|\mathbf{x}\rangle,
\end{equation*}
where $f(\mathbf{a})$ matches~(\ref{eq:obj_function}) term by term. Hence, each basis state is an eigenstate with an eigenvalue equal to the classical cost evaluated at the corresponding constellation vector. 
\end{IEEEproof}

\bibliographystyle{IEEEtran}
\bibliography{Ref_Quantum}

\end{document}